\documentclass[aps,prl,superscriptaddress,floatfix,amsmath,amssymb,amsfonts,reprint]{revtex4-2}

\usepackage[T1]{fontenc}
\usepackage{newtxtext} 
\usepackage{amsthm}
\usepackage{thmtools}
\usepackage{bm}        
\usepackage[dvipsnames]{xcolor}
\usepackage{enumerate}
\usepackage{graphicx}
\usepackage{microtype} 
\usepackage{multirow} 
\usepackage[normalem]{ulem}
\usepackage{xspace} 
\usepackage{float}
\usepackage{scalerel}
\usepackage{cancel}
\usepackage{braket}

\usepackage[colorlinks=true, linkcolor=blue, citecolor=blue, urlcolor=blue, bookmarksnumbered=true, pdfstartview=FitH]{hyperref}
\usepackage[all]{hypcap}
\usepackage{orcidlink}

\allowdisplaybreaks 

\usepackage{zref-clever} 
\zcsetup{nameinlink = false} 

\newcommand{\cref}[1]{\zcref{#1}}
\newcommand{\Cref}[1]{\zcref[S]{#1}} 

\zcRefTypeSetup{equation}{
  Name-sg = Equation,
  name-sg = Eq.,
  Name-pl = Equations,
  name-pl = Eqs.,
  refbounds = {(,,,)}, 
}

\zcRefTypeSetup{figure}{
  Name-sg = Figure,
  name-sg = Fig.,
  Name-pl = Figures,
  name-pl = Figs.,
}

\zcRefTypeSetup{section}{
  Name-sg = Section,
  name-sg = Sec.,
  Name-pl = Sections,
  name-pl = Secs.,
}

\zcRefTypeSetup{table}{
  Name-sg = Table,
  name-sg = Table,
  Name-pl = Tables,
  name-pl = Tables,
}

\zcRefTypeSetup{appendix}{
  Name-sg = Appendix,
  name-sg = App.,
  Name-pl = Appendices,
  name-pl = Apps.,
}

\newcommand{\mc}[1]{\ensuremath{\mathcal{#1}}}
\newcommand{\mr}[1]{\ensuremath{\mathrm{#1}}}

\newcommand{\la}{\ensuremath{\langle}}
\newcommand{\ra}{\ensuremath{\rangle}}

\newcommand{\matel}[3]{\ensuremath{\langle #1 | #2 | #3 \rangle}}

\newcommand{\mi}{\ensuremath{\mathrm{i}}} 
\newcommand{\me}{\ensuremath{\mathrm{e}}} 

\newcommand{\tvd}{{\scriptscriptstyle{\mathrm{TVD}}}}
\newcommand{\kl}{{\scriptscriptstyle{\mathrm{KL}}}}
\newcommand{\rel}{{\scriptscriptstyle{\mathrm{Rel}}}}
\newcommand{\pU}[1]{\ensuremath{p_U^{ #1 }}}

\makeatletter
\def\maketitle{
\@author@finish
\title@column\titleblock@produce
\suppressfloats[t]}
\makeatother

\newcommand{\PaperTitle}{Coherent-disorder-driven complexity transitions in a quantum-advantage architecture}

\begin{document}

\title{\PaperTitle}

\author{Sung-Bin B. Lee\,\orcidlink{0009-0009-2340-3682}}
\email{rqtoe@snu.ac.kr}
\affiliation{Department of Physics and Astronomy, Seoul National University, Seoul 08826, South Korea}
\affiliation{Center for Theoretical Physics, Seoul National University, Seoul 08826, South Korea}

\author{Chae-Yeun Park\,\orcidlink{0000-0003-4430-1000}}
\affiliation{School of Integrated Technology, Yonsei University, Seoul 03722, South Korea}
\affiliation{Department of Quantum Information, Yonsei University, Seoul 03722, South Korea}

\author{Changhun Oh\,\orcidlink{0000-0003-2002-1928}}
\affiliation{Department of Physics, Korea Advanced Institute of Science and Technology, Daejeon 34141, South Korea}

\author{Seung-Sup B.~Lee\,\orcidlink{0000-0003-0715-5964}}
\email{sslee@snu.ac.kr}
\affiliation{Department of Physics and Astronomy, Seoul National University, Seoul 08826, South Korea}
\affiliation{Center for Theoretical Physics, Seoul National University, Seoul 08826, South Korea}
\affiliation{Institute for Data Innovation in Science, Seoul National University, Seoul 08826, South Korea}

\date{\today}

\begin{abstract}
While decoherence is known to erode classical hardness in quantum random sampling, the impact of coherent spatial disorder remains an open question. We study a square-lattice instantaneous quantum polynomial-time (IQP) architecture subject to two-qubit gate-angle disorder and single-qubit dephasing using exact tensor-network simulations up to 576 qubits.  For finite systems without dephasing, increasing disorder drives two consecutive crossovers toward classical simulability: the output distribution first loses anticoncentration, and then the tensor-network simulation cost drops from exponential to polynomial as entanglement is suppressed. The finite-size scaling collapses are consistent with continuous transitions in the large-system limit. Dephasing further reduces the complexity. We characterize the computationally hard regime through scaling laws that provide quantitative error-budget bounds for realistic near-term devices.
\end{abstract}

\maketitle

\textit{Introduction}---%
Quantum random sampling is a leading route toward computational advantage of quantum machines over classical computers~\cite{Harrow2017, Boixo2018, Bouland2019, Hangleiter2023, Movassagh2023}.
In contrast to algorithms such as Shor's algorithm~\cite{Shor1997}, which are expected to require large-scale fault-tolerant hardware, sampling experiments are amenable to intermediate-scale noisy devices. A prominent example is the sampling of outputs from instantaneous quantum polynomial-time (IQP) circuits~\cite{Shepherd2009, Bremner2016, Bluvstein2024}, which consist of a few layers of diagonal qubit gates that mutually commute.

Despite the gate constraints and shallow circuit depth, classical simulation of general IQP circuits is widely believed to be computationally intractable~\cite{Bremner2010, Bremner2017, Hangleiter2023}. 
The standard sampling-hardness argument invokes two conjectures~\cite{Aaronson2011, Bouland2019, Bremner2016, BermejoVega2018, Park2025a}.
The first is that approximating output probabilities up to constant multiplicative error is $\#\textsf{P}$-hard for at least a constant fraction of circuit instances (average-case $\#\textsf{P}$-hardness).
It is motivated by the worst-case hardness: IQP output probabilities are expressed as complex-temperature Ising partition functions~\cite{Bremner2016, Fujii2017}, whose exact evaluation is $\#\textsf{P}$-hard in general~\cite{Jerrum1987, Jaeger1990}. Moreover, certain IQP circuits made of single- and two-qubit gates on a 2D planar graph can be used for universal quantum computation under postselection~\cite{Bremner2010}, which is expected to be classically unsimulable.
The second conjecture is the anticoncentration of the output distribution; the probability $p(x)$ of seeing a bitstring $x$ after a measurement is spread over many bitstrings.
Thus $p(x) \sim 2^{-n}$ for a non-negligible fraction of $x$'s, where $n$ is the number of qubits.
Given the two conjectures, an efficient classical sampler within
small total-variation distance is unlikely to exist, since it would
imply a collapse of the polynomial hierarchy (\textsf{PH}) to its third level via Stockmeyer's
approximate counting argument~\cite{Stockmeyer1983, Toda1991, Han1997, Aaronson2005}.

IQP circuits are attractive from the standpoint of experimental feasibility, as they can be implemented by initializing the qubits to a product state and then evolving them according to an Ising Hamiltonian for a constant time~\cite{Liu2025}.
Particularly, certain square-lattice IQP circuits generated by uniform nearest-neighbor Ising Hamiltonians are proposed as quantum-advantage architectures~\cite{Gao2017, BermejoVega2018, Haferkamp2020}, which can be realized by cold atoms in optical lattices.

However, it remains largely unexplored whether the quantum advantage of IQP architectures is robust against realistic imperfections.
While the effects of decoherence on IQP circuits have been studied extensively~\cite{Fujii2016, Bremner2017, Park2025, Rajakumar2025, Placidi2026}, considerably less is known about how coherent disorder---spatially inhomogeneous miscalibrations in gate parameters---affects their computational complexity. 
Such coherent errors are expected to become increasingly important as incoherent errors are reduced~\cite{Greenbaum2017, Bravyi2018, Hakkaku2021, Huang2019, Feng2016, Ahsan2022, Cai2020}.

In this work, we investigate how coherent disorder affects the practical simulability of Architecture I from Ref.~\cite{BermejoVega2018}, which is a square-lattice IQP circuit. Using exact tensor-network contractions to evaluate output probabilities, we demonstrate that two-qubit interaction disorder drives two crossovers, each undermining one of the two conjectural ingredients of sampling hardness: the first breaks anticoncentration, and the second makes typical output-probability evaluation tractable by suppressing tensor-network entanglement from volume-law to logarithmic scaling. Both the collision probability and the entanglement exhibit finite-size scaling (FSS) collapses, implying that these crossovers potentially sharpen into continuous transitions in the large-system limit.
In the End Matter, we show that introducing single-qubit dephasing further reduces the complexity, but differently from coherent disorder.

\begin{figure}[t]
    \includegraphics[width=\linewidth]{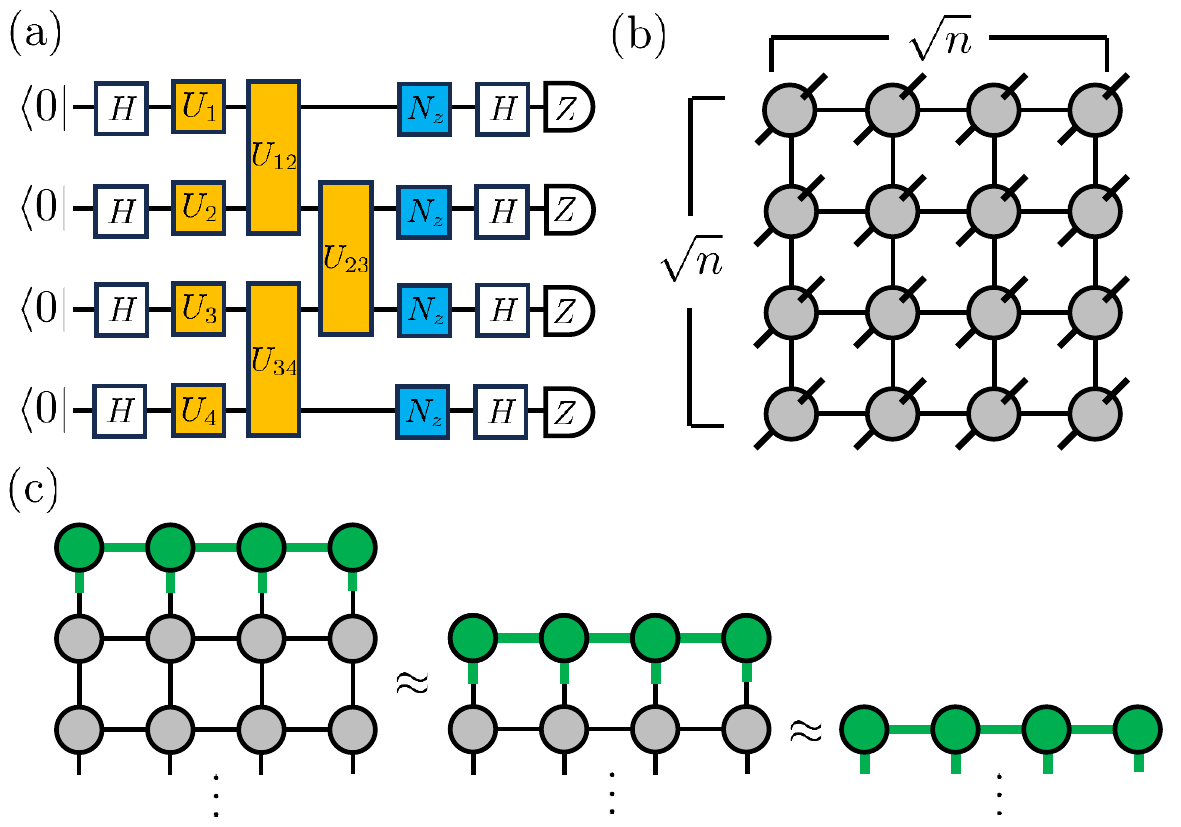}
    \caption{
    (a) An instantaneous quantum polynomial-time (IQP) circuit consists of single- and two-qubit gates ($U_i$, $U_{ij}$) diagonal in the computational basis, between the two layers of Hadamard ($H$) gates.
    We also consider single-qubit dephasing noise channels ($N_z$) of strength $\varepsilon$.
    (b) The density operator for the circuit on a $\sqrt{n} \times \sqrt{n}$ square lattice can be expressed as a projected entangled-pair operator (PEPO) on the same lattice, which has two physical legs per site and bond dimensions up to four.
    When there is no dephasing ($\varepsilon = 0$), we use a projected entangled-pair state (PEPS) that has one physical leg per site and bond dimensions up to two.
    (c) A square-lattice tensor network (gray), given by projecting the physical legs of PEPS or PEPO onto a bitstring state, is contracted by using a boundary matrix product state (BMPS) (green).
    }
\label{fig:pepo_representation}
    \vspace{-1em}
\end{figure}

\textit{Setup and method}---%
\Cref{fig:pepo_representation} describes our system and method. We consider qubits on a square lattice of size $\sqrt{n} \times \sqrt{n}$, where $\sqrt{n}$ is even and the qubits can interact only with their nearest neighbors. Every qubit is first prepared as $(\ket{0} + \ket{1})/\sqrt{2}$. Then we apply $U := \prod_{\la i j \ra} U_{ij} \prod_{i} U_i$, where $U_{ij} = \exp(\mi J_{ij} Z_i Z_j)$ and $U_i = \exp( - \mi h_i Z_i)$ are, respectively, two- and single-qubit gates generated by Pauli-$Z$ operators. $U$ can be viewed as real-time evolution for a constant time under a nearest-neighbor Ising Hamiltonian $\mc{H} = - \sum_{\la i j \ra} J_{ij} Z_i Z_j + \sum_i h_i Z_i$.
We draw each ``magnetic field'' parameter $h_i$ independently and uniformly from $\{\pi/4,3\pi/8\}$ with equal probabilities, and each coupling as $J_{ij}=(\pi/4)(1+\sigma\eta_{ij})$, where $\eta_{ij}$ is a standard normal random variable $\mc{N}(0,1)$.
On top of this, every qubit experiences a single-qubit dephasing channel, $N_z: \rho_i \to (1 - \varepsilon/2) \rho_i + (\varepsilon/2) Z_i \rho_i Z_i$. The gates are diagonal in the computational basis, and the dephasing channel is diagonal in the corresponding Pauli-transfer representation; hence all operations commute. Finally, we measure every qubit in the $(\ket{0} \pm \ket{1})/\sqrt{2}$ basis, which yields a bitstring $x \in \{0, 1\}^{n}$ with probability $\pU{\sigma\varepsilon}(x)$.

When $\sigma=\varepsilon=0$, all couplings are uniform, $J_{ij}=\pi/4$, and the circuit reduces to Architecture I from Ref.~\cite{BermejoVega2018}. The randomness in the $h_i$ values is part of that ideal architecture, not the coherent disorder studied here. These nonzero local fields are essential for moving beyond the Pfaffian-solvable zero-field planar Ising case~\cite{Kasteleyn1961,Fisher1961,Fujii2017}. In this work, disorder refers only to the quenched two-qubit angle fluctuations $\eta_{ij}$, whose strength is $\sigma$; $\varepsilon$ independently denotes the dephasing strength.

Because the gates commute, the dephasing-free ($\varepsilon = 0$) circuit state can be represented as a PEPS, whose bond dimensions are $2$. Such small bond dimensions distinguish IQP from universal random circuits, whose PEPS representation has bond dimensions exponential in circuit depth~\cite{Lee2025}. When $0 < \varepsilon < 1$, we describe a mixed state of the circuit as a PEPO [see \cref{fig:pepo_representation}(b)], whose bond dimensions are the squares of those of the PEPS. Then the probability $\pU{\sigma\varepsilon}(x)$ is given by the contraction of the two-dimensional tensor network constructed by projecting the physical legs of the PEPS or PEPO onto $\ket{x}$.
For the contraction, we use the boundary matrix product state (BMPS) method~\cite{Verstraete2004,Murg2007} [see \cref{fig:pepo_representation}(c)], a standard technique for contracting finite 2D tensor networks. The contraction algorithm initializes the BMPS using the first row of tensors. Then the BMPS absorbs the next row at each iteration, while the bonds are not truncated; the contraction is exact up to double precision.

In this work, we study various quantities based on $\pU{\sigma\varepsilon}(x)$: the rescaled collision probability $\tilde{Z} = 2^n \sum_x \pU{\sigma \varepsilon} (x)^2$, the total variation distance (TVD) $\delta_\tvd (\pU{0 0}, \pU{\sigma \varepsilon}) = \frac{1}{2} \sum_x | \pU{0 0} (x) - \pU{\sigma \varepsilon} (x) |$, the Kullback--Leibler divergence (KLD) $\delta_\kl (\pU{0 0} || \pU{\sigma \varepsilon}) = \sum_x \pU{0 0} (x) \ln [\pU{0 0} (x) / \pU{\sigma \varepsilon} (x)]$, and the relative error $\delta_\rel (\pU{0 0} || \pU{\sigma \varepsilon}) = 2^{-n} \sum_x |\pU{\sigma \varepsilon} (x) - \pU{0 0}(x)| / \pU{0 0}(x)$.  Their definitions involve summation over all possible bitstrings $x$, which is infeasible due to the exponentially large $| \{ x \}| = 2^{n}$. Instead, we estimate these quantities by averaging over samples, namely replacing $\sum_x \to (2^n / |\mc{X}|) \sum_{x \in \mc{X}}$. Here $\mc{X}$ is a set of uniformly chosen bitstring samples, with $| \mc{X} | = 2^{12} = 4{,}096$. Furthermore, most of the data shown in this work are obtained by averaging over 100 circuit instances. An exception is \cref{fig:peps_schmidt_decay}, where the entanglement spectrum of each circuit realization is shown as a separate line.

\textit{Ideal-case hardness---}%
At the tensor-network level, the difficulty of evaluating the ideal output probability $\pU{00}(x)$ is reflected in the volume-law entanglement of the BMPS (cf.~\cref{fig:entanglements}), together with the nearly flat entanglement spectrum (cf.~\cref{fig:peps_schmidt_decay}). This implies that exact tensor-network contractions require exponentially large BMPS bond dimensions. 
We find that the probability density function of $\pU{00}(x)$ follows the Porter--Thomas distribution (PTD)~\cite{Porter1956} (see \cref{fig:PDF}), which is consistent with the observation in Ref.~\cite{BermejoVega2018}.
In the PTD, $\pU{00}(x)$ is broadly spread over bitstrings, i.e., anticoncentrated.
For the rest of the main text, we investigate how the statistical and entanglement properties evolve as $\sigma$ increases from $0$, while $\varepsilon$ is kept $0$. We discuss the effect of finite $\varepsilon$ in the End Matter.

\begin{figure*}
    \includegraphics[width=\textwidth]{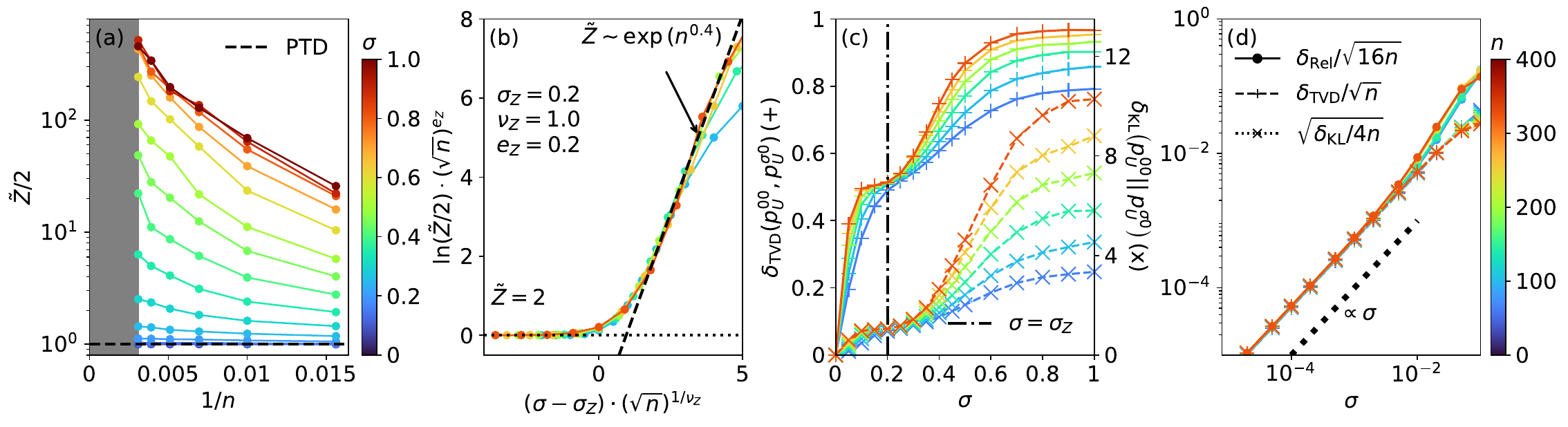}
    \caption{
    (a) The rescaled collision probability $\tilde{Z}$ as a function of $1/n$, where line colors indicate disorder strength $\sigma$.
    The dotted line marks the PTD value $\tilde{Z} = 2$.
    (b) Finite-size scaling (FSS) collapse of $\tilde{Z}(n,\sigma)$. The dotted and dashed lines represent the small- and large-argument asymptotes.
    (c, d) Statistical distances between the bitstring probability with disorder and no dephasing, $\pU{\sigma 0} (x)$, and the ideal case, $\pU{0 0} (x)$.
    (c) The TVD $\delta_\tvd$ ($+$'s) and KLD $\delta_\kl$ ($\times$'s) as functions of $\sigma$ on a linear scale.
    The vertical dash-dotted line marks $\sigma = \sigma_{Z} = 0.2$.
    (d) Rescaled $\delta_\tvd$, $\delta_\kl$, and relative error $\delta_\rel$ as functions of $\sigma$ on a log scale.
    In panels (b)--(d), colors encode the number of qubits $n$ on a linear scale.
    }
    \label{fig:distance_scaling}
    \vspace{-1em}
\end{figure*}

\textit{Output statistics---}%
We first investigate the crossover at a smaller $\sigma$, which appears in the distribution of $\pU{\sigma 0}(x)$. As a measure of lack of anticoncentration, we compute the rescaled collision probability $\tilde{Z}$, where the original collision probability $2^{-n} \tilde{Z} = \sum_x \pU{\sigma 0} (x)^2$ represents the probability that two independent measurements yield the same output. \Cref{fig:distance_scaling}(a) shows that $\tilde{Z}$ grows rapidly as $n$ increases for larger $\sigma$, while $\tilde{Z}$ remains bounded for smaller $\sigma$.

To be more quantitative, we perform the finite-size scaling (FSS)~\cite{Fisher1972} of $\tilde{Z}(n, \sigma)$ using an ansatz $\ln (\tilde{Z}/2) = (\sqrt{n})^{-e_Z} \mc{F}_Z [(\sigma - \sigma_Z) (\sqrt{n})^{1/\nu_Z}]$, inspired by the FSS for second-order phase transitions. The data points for $\sigma \lesssim 0.4$ collapse onto a single curve $\mc{F}_Z$ for $(\sigma_Z, \nu_Z, e_Z) \simeq (0.2, 1.0, 0.2)$, as shown in \cref{fig:distance_scaling}(b). $\mc{F}_Z(x)$ smoothly connects between a horizontal line at $0$ for $x \lesssim -0.5$ and a linearly increasing line for $x \gtrsim 2$, which implies $\tilde{Z} \to 2$ (the PTD value, anticoncentrated) for $\sigma < \sigma_Z$ and $\tilde{Z} \sim \exp [n^{(1/\nu_Z-e_Z)/2}] = \exp(n^{0.4})$ (non-anticoncentrated) for $\sigma > \sigma_Z$. The latter is consistent with, and stronger than, the lower-bound form for non-anticoncentrated cases, $\tilde Z\geq \exp(n^c)$ with $c$ independent of $n$, proposed in Ref.~\cite{Dalzell2022}.
Assuming the scaling form persists to larger systems, the crossover sharpens into a transition at $\sigma_Z$ between anticoncentrated and non-anticoncentrated regimes.

A similar crossover is also captured by the statistical distances between the probability under disorder $\{ \pU{\sigma 0} (x) \}_x$ and the ideal-case probability $\{ \pU{0 0} (x) \}_x$.
\Cref{fig:distance_scaling}(c) depicts $\delta_\tvd (\pU{0 0}, \pU{\sigma 0})$ and $\delta_\kl (\pU{0 0} || \pU{\sigma 0})$, which both increase as either $\sigma$ or $n$ grows.
Notably, the curves for different $n$'s intersect near $\sigma=\sigma_Z$, indicating approximate $n$-independence near the crossover.
Within the FSS interpretation, the crossings near $\sigma_Z$ are consistent with a growing length scale $\xi_Z \sim |\sigma-\sigma_Z|^{-\nu_Z}$ with $\nu_Z \simeq 1$.

We remark that in the low-$\sigma$ regime the statistical distances exhibit characteristic scalings, $\delta_\rel \propto \delta_\tvd \simeq \sqrt{\delta_\kl} / 2 \propto \sqrt{n} \sigma$, as shown in \cref{fig:distance_scaling}(d).
Interestingly, the relation $\delta_\kl \simeq 4 \delta_\tvd^2$ in our data resembles the bound given by the Pinsker inequality, $\delta_\tvd \leq \sqrt{\delta_\kl / 2}$, up to a factor of $2$.

\begin{figure}
    \includegraphics[width=\linewidth]{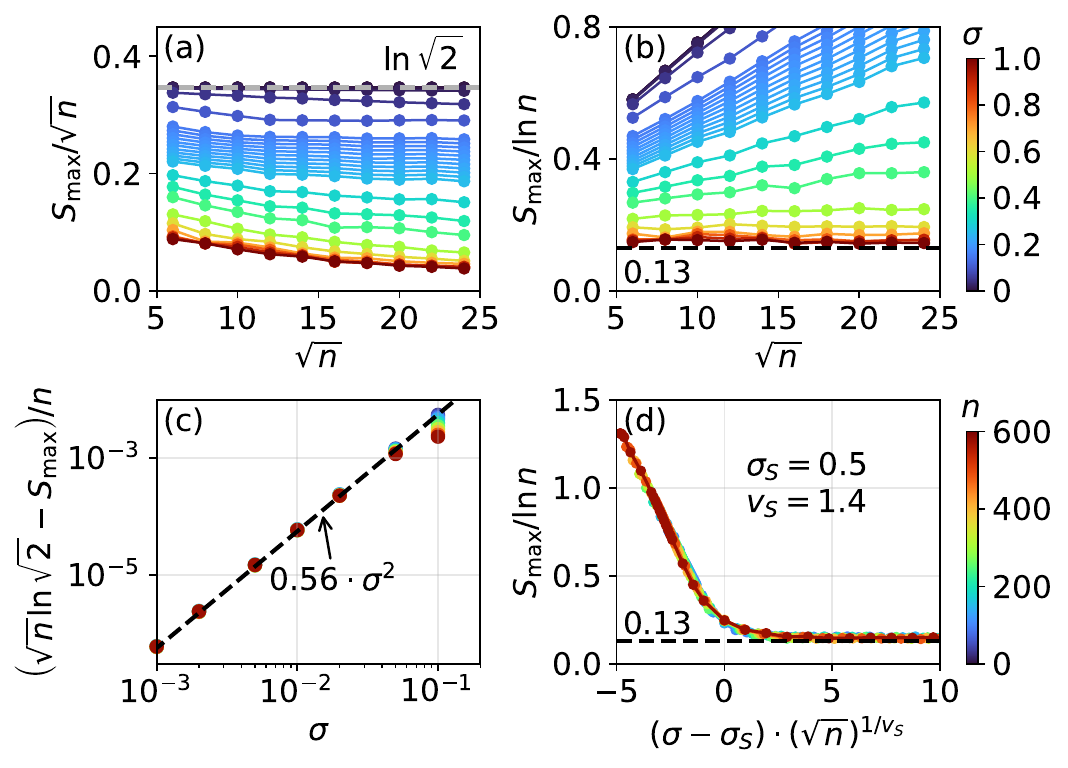}
    \caption{
    (a), (b) Maximum BMPS entanglement entropy $S_{\max}$ encountered during the tensor-network contraction for computing $\pU{\sigma0}(0)$;
    colors encode $\sigma$.
    The dashed lines in panels (a) and (b) mark, respectively, $S_{\max} = \sqrt{n} \ln \sqrt{2}$ and $S_{\max} = 0.13 \ln n$.
    (c) Small-$\sigma$ decrease of $S_{\max}$ from the ideal-case value, $(\sqrt n\ln\sqrt2-S_{\max})/n$.
    The dashed line is a fit to data for $\sigma\leq0.02$.
    (d) FSS collapse of $S_{\max}(n,\sigma)$.
    In panels (c) and (d), colors encode $n$.
    }
    \label{fig:entanglements}
    \vspace{-1em}
\end{figure}

\textit{Entanglement of BMPS---}%
We now turn to the entanglement of the BMPS, which determines the computational cost of our tensor-network simulations. The largest entanglement is typically encountered near the center of the BMPS obtained after contracting half of the rows of the square-lattice tensor network.
The corresponding maximum entanglement entropy $S_{\max}$ is shown in \cref{fig:entanglements}(a,b) as a function of system size $n$ and disorder strength $\sigma$. In the ideal case $\sigma=0$, we obtain $S_\mathrm{max}=\sqrt{n}\ln\sqrt{2}$, corresponding to the maximal possible entanglement of a BMPS of length $\sqrt{n}$, whose ``physical'' legs (which are actually the vertical bonds of PEPS) are $2$-dimensional. This volume-law entanglement requires an exponentially large bond dimension, $\chi=2^{\sqrt{n}/2}$, and therefore leads to the exponential contraction cost of $O(n\chi^4)=O(n 4^{\sqrt{n}})$.
As $\sigma$ is slightly increased, the entanglement decreases quadratically, $\sqrt{n} \ln \sqrt{2} - S_{\max} \propto n \sigma^2$; see \cref{fig:entanglements}(c). This dependence mirrors the small-$\sigma$ behavior of the KLD, $\delta_\kl \propto n\sigma^2$.

\begin{figure}
    \includegraphics[width=\linewidth]{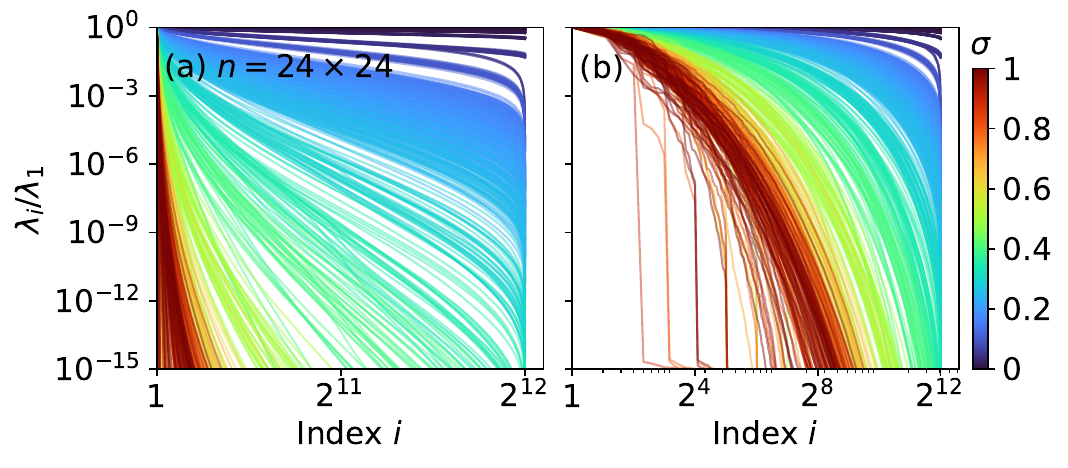}
    \caption{
    BMPS singular-value spectra at the bond where $S_{\max}$ is attained, plotted on (a) log-linear and (b) log-log scales. The singular values $\{\lambda_i\}_{i=1}^{2^{\sqrt n/2}}$ are sorted in descending order and rescaled by the largest value $\lambda_1$. Each line represents one circuit instance, and colors encode the disorder strength $\sigma$. For $\sigma>\sigma_S\simeq0.5$, the spectra decay exponentially, with only a small fraction of singular values remaining above the double-precision floor.
    }
    \label{fig:peps_schmidt_decay}
    \vspace{-1em}
\end{figure}

For larger $\sigma$, the entanglement undergoes a crossover from the volume-law to the logarithmic scaling of $S_{\max}$.
\Cref{fig:entanglements}(d) shows that the data for $S_{\max}(n,\sigma)$ collapse under the FSS ansatz $S_{\max} = (\ln n ) \mc{F}_S [(\sigma - \sigma_S) (\sqrt{n})^{1/\nu_S}]$ for $(\sigma_S, \nu_S) \simeq (0.5, 1.4)$. (Meanwhile, only the data points for $\sigma \lesssim 0.4$ show a collapse onto $\mc{F}_Z$ in \cref{fig:distance_scaling}(b).) The left branch of the scaling curve $\mc{F}_S (x \lesssim -2)$, on which the points of $S_{\max} = \sqrt{n} \ln \sqrt{2}$ for $\sigma = 0$ lie, is not a straight line.
By contrast, the right branch is flat, $\mc{F}_S (x \gtrsim 4) \simeq 0.13$.
Together with the exponentially decaying entanglement spectra for $\sigma>\sigma_S$ in \cref{fig:peps_schmidt_decay}, this suggests that the BMPS can be truncated at polynomial bond dimension, leading to polynomial contraction cost.
Similarly to the output-statistics crossover at $\sigma=\sigma_Z$, we expect the entanglement crossover near $\sigma=\sigma_S$ to sharpen into a transition from exponential to polynomial tensor-network contraction cost.

\begin{figure}
    \includegraphics[width=\linewidth]{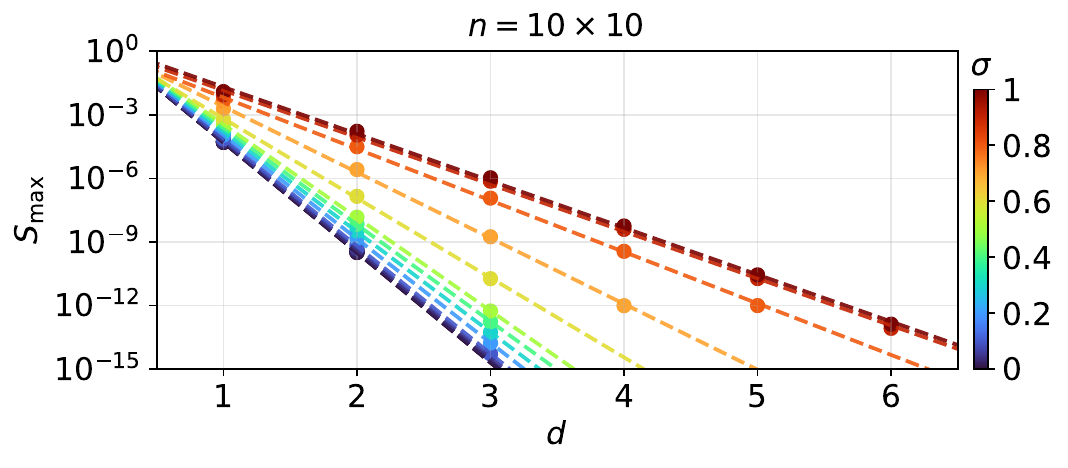}
    \caption{
    The maximum BMPS entanglement entropy $S_{\max}$ encountered when contracting the tensor network for imaginary-time evolution $\mc{U}=\me^{-d\mc{H}}$, where $\mc{H}$ is the Ising Hamiltonian that generates the IQP architecture. Colors encode $\sigma$ on a linear scale.
    }
    \label{fig:imaginary}
    \vspace{-1em}
\end{figure}

\textit{Imaginary-time evolution---}%
The amplitude $\bra{x} H^{\otimes n} U H^{\otimes n}$ $\ket{0}$ associated with a measurement outcome $x$ for a dephasing-free ($\varepsilon = 0$) IQP circuit is given by an imaginary-temperature Ising partition function with additional local fields acting on sites for which $x_i=1$~\cite{Bremner2016, Fujii2017}.
Given this, the disorder-induced reduction of complexity may appear to conflict with spin-glass intuition: in classical Ising models, disorder is often the source of computational hardness~\cite{Barahona1982, Istrail2000}. The resolution is that the IQP sampler involves real-time evolution, whereas classical spin-glass hardness concerns real Boltzmann weights. To make this distinction explicit, we simulate imaginary-time evolution under the same Ising Hamiltonian, $\mc{U}=\me^{-d\mc{H}}$, for which the bitstring amplitude $\matel{0}{H^{\otimes n}\mc{U}H^{\otimes n}}{0}$ yields the real-temperature partition function for $\mc{H}$ at temperature $1/d$.

\Cref{fig:imaginary} shows that $S_{\max}$ for the imaginary-time simulations decays exponentially with $d$. This reflects the fact that $\mc{H}$ is weakly frustrated in the relevant small-disorder regime: the fields $h_i$ are positive and the couplings $J_{ij}$ are mostly positive for small $\sigma$. Thus $\mc{U}$ rapidly becomes dominated by one or a few computational-basis product states. The decay rate of $S_{\max}$ decreases as $\sigma$ grows, because disorder in $J_{ij}$ introduces frustration. Nevertheless, the values of $S_{\max}$ remain exponentially smaller than their real-time counterparts for all $\sigma\in[0,1]$, consistent with the relative ease of imaginary-time tensor-network simulations compared to real-time ones~\cite{Grundner2024, Cao2024}.

\textit{Effect of dephasing---}%
Dephasing noise further reduces classical simulation complexity, but its effect differs qualitatively from that of coherent disorder. While both dephasing and disorder strongly suppress $S_{\max}$, dephasing decreases the collision probability, opposite to the effect of disorder. This difference arises from their respective effects on the probability density function (PDF): 
disorder broadens the PDF by producing both unusually high-probability and low-probability bitstrings, whereas dephasing drives the PDF toward a sharp peak at the uniform value $2^n p=1$.
In the computationally hard regime ($\sigma, \varepsilon \leq 0.01$), the squared relative error, squared TVD, and entanglement reduction scale as $\varepsilon^2$, whereas the KLD scales as $\varepsilon^{1.69}$. For a detailed discussion of these effects, see the End Matter.

\textit{Conclusion}---%
In summary, we have shown that coherent spatial disorder degrades the known hardness signatures and practical tensor-network hardness of a square-lattice IQP architecture. Using exact tensor-network simulations, we identified two distinct disorder-driven crossovers: loss of anticoncentration near $\sigma=\sigma_Z\simeq0.2$ and suppression of BMPS entanglement near $\sigma=\sigma_S\simeq0.5$. Our FSS analyses suggest that these crossovers sharpen in the large-$n$ limit into, respectively, a transition from anticoncentration to non-anticoncentration and a transition from exponential to polynomial tensor-network contraction cost.
In this regard, our model provides a benchmark for distinguishing the complexity mechanisms of IQP circuits and classical spin glasses, as well as those of real- versus imaginary-time tensor-network simulations.

Furthermore, this vulnerability to disorder distinguishes IQP circuits from deep universal random circuits. In sufficiently deep non-commuting random circuits, the circuit ensemble can approximate low-order moments of the Haar measure, so changing generic gate angles typically leaves the circuit in a Haar-like pseudorandom regime~\cite{Harrow2009,Brandao2016,Harrow2023}. In contrast, the restricted commuting structure of IQP circuits prevents such universal scrambling; spatial disorder instead disrupts the interference structure underlying anticoncentration and large BMPS entanglement.

\begin{acknowledgments}
We thank Byeongseon Go, Naoki Kawashima, Su-un Lee, Sheng-Hsuan Lin, Tsuyoshi Okubo, Feng-Feng Song, Simon Trebst, Xhek Turkeshi, Frank Verstraete, and Pan Zhang for fruitful discussions.
S.-B.B.L.~and S.-S.B.L.~were supported by the National Research Foundation of Korea (NRF) grants funded by the Korean government (MSIT: No.~RS-2023-00214464, No.~RS-2024-00442710, No.~RS-2023-00258359, No.~RS-2023-NR119928, No.~RS-2024-00413957; MEST: No.~2019R1A6A1A10073437), the Global-LAMP Program funded by the Ministry of Education (No.~RS-2023-00301976), and Samsung Electronics Co., Ltd.~(No.~IO220817-02066-01).
C.Y.P.~was supported by the Yonsei University Research Fund of 2025-22-0436, and the NRF grants funded by the Korean government (MSIT) (No.~RS-2025-16066935, RS-2025-02316431, and  RS-2025-18362970).
C.O.~was supported by the National Research Foundation of Korea Grants (No.~RS-2024-00431768 and No.~RS-2025-00515456) funded by the Korean government (Ministry of Science and ICT (MSIT)) and the Institute of Information \& Communications Technology Planning \& Evaluation (IITP) Grants funded by the Korean government (MSIT) (No.~RS-2024-00437284, No.~IITP-2025-RS-2025-02283189 and No.~IITP-2025-RS-2025-02263264).
\end{acknowledgments}

%

\clearpage

\onecolumngrid
\begin{center}
    \fontsize{12}{16}\selectfont
    \textbf{End Matter}
\end{center}%
\twocolumngrid

\begin{figure}
    \includegraphics[width=\linewidth]{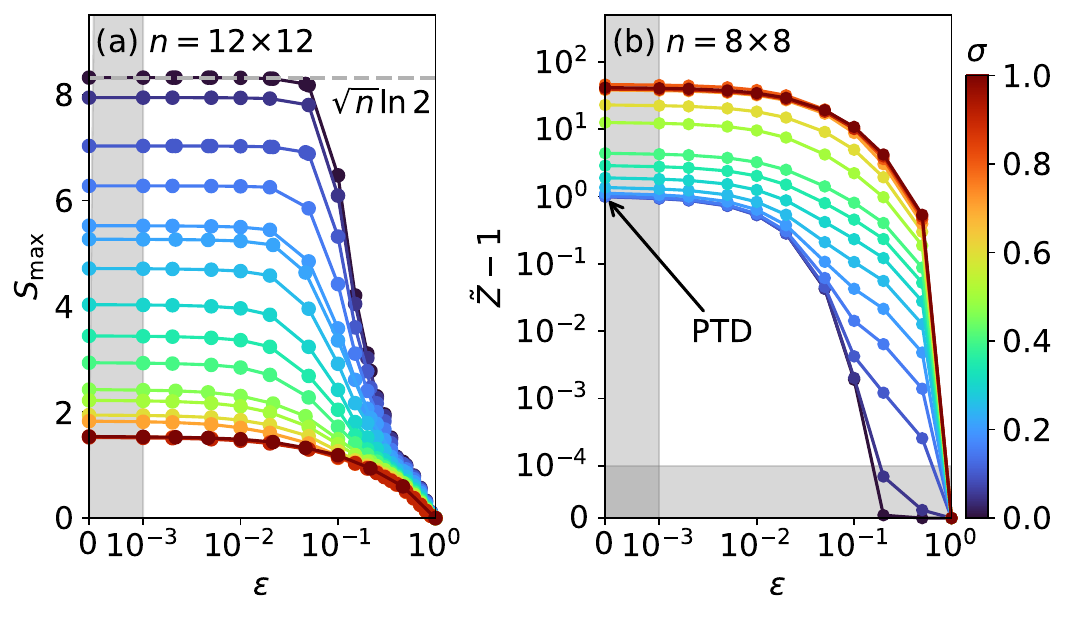}
    \caption{
    (a) $S_{\max}$ and (b) $\tilde{Z}-1$ as functions of $\varepsilon$. Colors encode $\sigma$ on a linear scale; $n$ is indicated in each panel.
    For both panels, $\varepsilon$ is shown on a logarithmic scale for $\varepsilon > 10^{-3}$ and on a linear scale for $\varepsilon < 10^{-3}$ (gray shades).
    The dashed line in panel (a) marks the ideal-case $(\sigma=\varepsilon=0)$ entanglement.
    In panel (b), $\tilde{Z}-1$ is shown on a logarithmic scale for $\tilde{Z}-1 > 10^{-4}$ and on a linear scale for $\tilde{Z}-1 < 10^{-4}$ (gray shade).
    }
    \label{fig:pepo_entanglement_and_decay}
    \vspace{-1em}
\end{figure}

\begin{figure}
    \includegraphics[width=\linewidth]{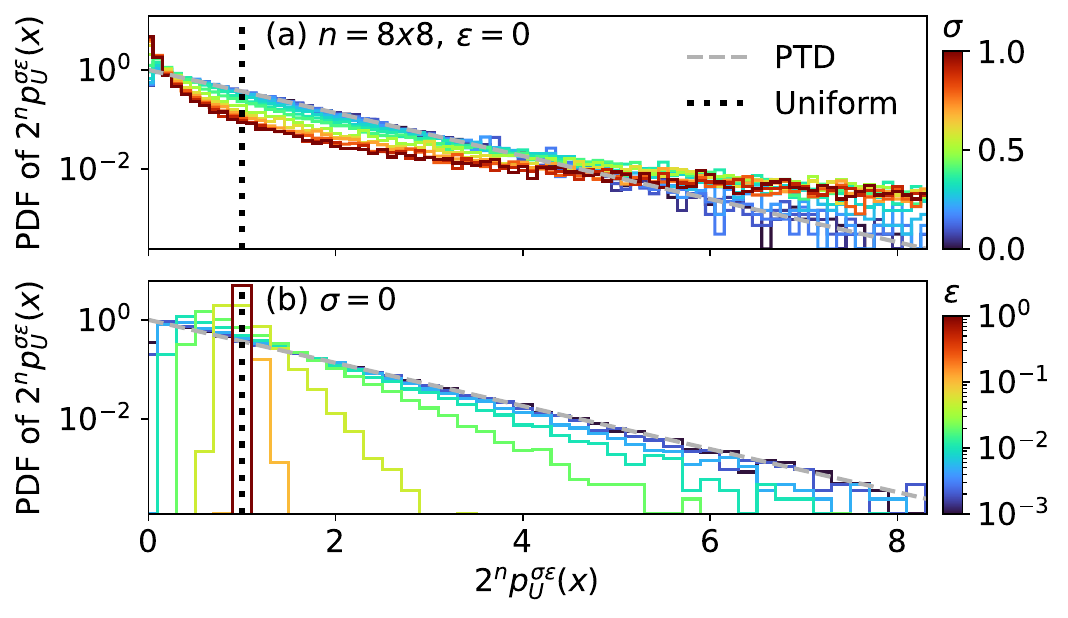}
    \caption{
    Probability density function (PDF) of the rescaled bitstring probability $y=2^n\pU{\sigma\varepsilon}(x)$ for $n=8\times8$, estimated from $2^{12}$ uniformly chosen bitstrings. Colors encode (a) $\sigma$ on a linear scale while $\varepsilon=0$, and (b) $\varepsilon$ on a logarithmic scale while $\sigma=0$. The dashed line indicates the PTD, $\mathrm{Pr}(y)=e^{-y}$, while the dotted line indicates the uniform distribution, $p(x)=2^{-n}$ for all $x$.
    }
    \label{fig:PDF}
    \vspace{-1em}
\end{figure}

\textit{Entanglement and collision probability under dephasing---}%
\Cref{fig:pepo_entanglement_and_decay}(a) shows that $S_{\max}$ decreases as either $\sigma$ or $\varepsilon$ grows.
Note that here we contract square-lattice tensor networks with bond dimensions up to 4, which originate from PEPOs that describe dephased mixed states. Hence the maximal value of $S_{\max}$ is $\sqrt{n} \ln 2$ here, in contrast to $\sqrt{n} \ln \sqrt{2}$ from the PEPS-based simulations [cf.~\cref{fig:entanglements}].

By contrast, larger $\varepsilon$ decreases $\tilde{Z}$, while larger $\sigma$ increases $\tilde{Z}$, as shown in \cref{fig:pepo_entanglement_and_decay}(b).
This contrast arises because the PDF of the bitstring probabilities changes differently under disorder and dephasing, as shown in \cref{fig:PDF}.
In the ideal case ($\sigma = \varepsilon = 0$), the PDF follows the PTD.
Without dephasing ($\varepsilon = 0$), larger $\sigma$ suppresses the PDF for $0.3 \lesssim 2^n \pU{\sigma 0}(x) \lesssim 4.8$, while enhancing it outside this interval.
This broadening is compatible with approximate tensor-network strategies, such as bond truncation, whose effects are most visible in the low-probability part of the output PDF; see the PDFs obtained from truncated PEPS in Fig.~7 of Ref.~\cite{Lee2025}.
On the other hand, without disorder ($\sigma = 0$), the PDF collapses toward $2^n \pU{0\varepsilon}(x) = 1$ as $\varepsilon$ increases.

\begin{figure}
    \includegraphics[width=\linewidth]{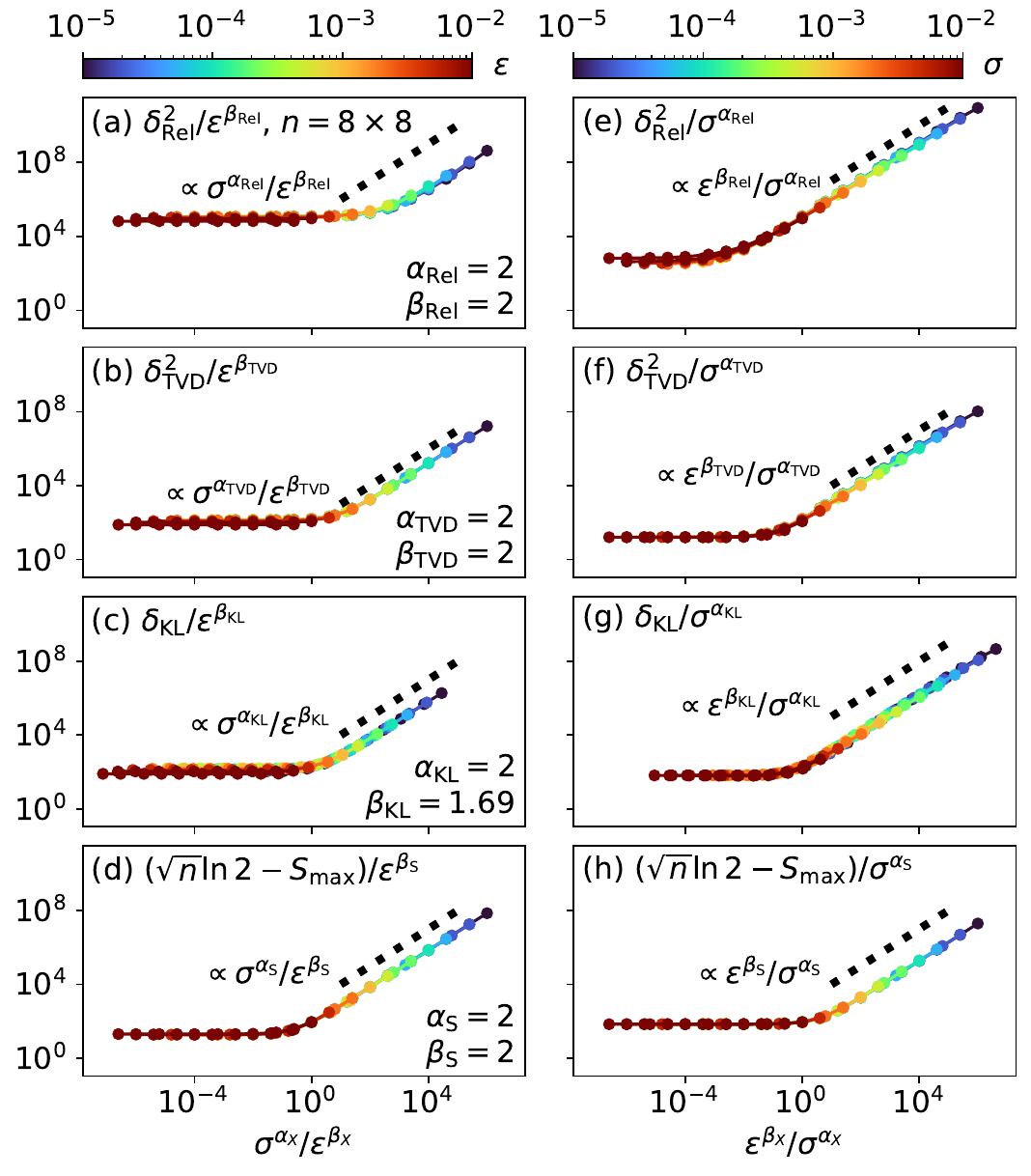}
    \caption{
    (a),(b) Scaling collapses of the squared relative error and squared TVD for small disorder and dephasing, $\sigma, \varepsilon \leq 0.01$, plotted as $\delta_X^2/\varepsilon^{\beta_X}$ versus $\sigma^{\alpha_X}/\varepsilon^{\beta_X}$ for $X=\mr{Rel},\mr{TVD}$.
    (c),(d) Analogous collapses for the KLD, $\delta_\kl$ ($X=\mr{KL}$), and the entanglement reduction, $\sqrt{n} \ln 2  - S_{\max}$ ($X=S$).
    (e)--(h) The same data represented with $\sigma$ and $\varepsilon$ interchanged, so the horizontal axis is the inverse of that in the corresponding left-column panel.
    All panels use $n=8\times8$.
    }
    \label{fig:pepo_error_scaling}
    \vspace{-1em}
\end{figure}

\textit{Statistical distances and entanglement reduction for small $\sigma$ and $\varepsilon$---}%
In \cref{fig:pepo_error_scaling}(a)--(d), we show scaling collapses for the small-$\sigma$ and small-$\varepsilon$ regime. For each quantity $X$, the data collapse when divided by $\varepsilon^{\beta_X}$ and plotted against the ratio $\sigma^{\alpha_X}/\varepsilon^{\beta_X}$, where $\alpha_X$ and $\beta_X$ are the disorder and dephasing exponents, respectively. Panels (e)--(h) show the same data with $\sigma$ and $\varepsilon$ interchanged.
These collapses indicate that each plotted quantity scales as $\varepsilon^{\beta_X}$ when $\sigma^{\alpha_X}/\varepsilon^{\beta_X}$ is small and as $\sigma^{\alpha_X}$ when this ratio is large.
The crossover of $\delta_\rel^2$ occurs near $\sigma^{\alpha_X}/\varepsilon^{\beta_X}=10^2$, whereas the other crossovers occur near $\sigma^{\alpha_X}/\varepsilon^{\beta_X}=1$.
All fitted exponents are close to $2$ except for $\beta_\kl=1.69$, indicating that the KLD responds differently to dephasing than the other quantities.

\end{document}